# Effects of Steric Factors on Molecular Doping to MoS$_2$.

Serrae N. Reed[1,2], Yifeng Chen[3], Milad Yarali[1,2], David J. Charboneau[4], Julia B. Curley[4], Nilay Hazari[4], Su Ying Quek[3,5,6,7], Judy J. Cha[1,2]

1. Department of Mechanical Engineering and Materials Science, Yale University, New Haven, CT 06511, USA
2. Energy Sciences Institute, Yale West Campus, West Haven, CT 06516, USA
3. Department of Physics, National University of Singapore, 117551, Singapore
4. Department of Chemistry, Yale University, New Haven, CT 06511, USA
5. Centre for Advanced 2D Materials, National University of Singapore, Block S14, Level 6, 6 Science Drive 2, 117546, Singapore
6. NUS Graduate School, Integrative Sciences and Engineering Programme, National University of Singapore, Singapore 117456, Singapore
7. Department of Materials Science and Engineering, National University of Singapore, Singapore 117575, Singapore

**Surface functionalization of two-dimensional (2D) materials with organic electron donors (OEDs) is a powerful method to modulate the electronic properties of the material. However, our fundamental understanding of the doping mechanism is largely limited to the categorization of molecular dopants as n- or p-type based on the relative position of the molecule's redox potential in relation to the Fermi level of the 2D host. Our limited knowledge about the impact of factors other than the redox properties of the molecule on doping makes it challenging to controllably use molecules to dope 2D materials and design new OEDs. Here, we functionalize monolayer MoS$_2$ using two molecular dopants, Me- and $^t$Bu-OED, which have the same redox potential but different steric properties to probe the effects of molecular size on the doping level of MoS$_2$. We show that, for the same functionalization conditions, the doping powers of Me- and $^t$Bu-OED are 0.22-0.44 and 0.11 electrons per molecule, respectively, demonstrating that the steric properties of the molecule critically affect doping levels. Using the stronger dopant, Me-OED, a carrier density of $1.10 \pm 0.37 \times 10^{14}$ cm$^{-2}$ is achieved in MoS$_2$, the highest doping level to date for**

MoS$_2$ using surface functionalization. Overall, we establish that tuning of the steric properties of the dopant is essential in the rational design of molecular dopants.

**Introduction**

The surface functionalization of 2D materials with organic molecules is a powerful tool to modulate the properties of 2D materials, in part due to our ability to synthesize molecular dopants with specific structures and functionalities. Since the seminal surface functionalization of graphene with organic electron donors (OEDs) in 2007, the properties of a wide variety of 2D materials have been modified using this method.[1-22] Particularly noteworthy results include the reversible doping of MoS$_2$ using benzyl viologen[7] and the extremely high carrier density of 5.8 ± 1.9 × 10$^{13}$ cm$^{-2}$ achieved in MoS$_2$ via surface functionalization with an organic dopant based on 2,2'-bipyridine (DMAP-OED).[3] Generally, the amount of charge doping from OEDs is predicted based on the relative position of the molecule's redox potential in relation to the Fermi level of the 2D host.[5, 23] However, other factors such as the steric properties of the molecule, the nature of the interaction between the molecule and the 2D material, and interactions between molecules on the 2D material are also likely to impact the overall charge donation. For example, we found that the doping power of DMAP-OED to monolayer MoS$_2$ can vary as a function of molecular surface coverage.[3] Thus, understanding the effects of molecular structure, surface coverage, and molecule-molecule interactions on charge doping is essential for the rational design of molecular dopants. Nevertheless, systematic experimental studies that examine the effects of molecular properties beyond redox potential on molecular charge doping are currently lacking.

Here, we examine the effects of the steric properties of molecular dopants on charge doping to monolayer MoS$_2$ flakes using two OEDs with the same redox potential but with substituents of different size: methyl (Me) and tert-Butyl ($^t$Bu) groups. We quantify the doping powers of the two OEDs by measuring the change in carrier density of MoS$_2$ field-effect transistors (FETs) before and after functionalization and estimating the number of molecules that functionalize MoS$_2$ using atomic force microscopy (AFM). For the same functionalization conditions, we show that the doping powers of Me- and $^t$Bu-OED are 0.22-0.44 and 0.11 electrons per molecule, respectively. Thus, despite having the same redox potential, the two OEDs donate different amounts of electrons to MoS$_2$ due to differences in their size, which affect their interaction with MoS$_2$. Density functional theory (DFT) calculations are performed for the two OEDs at different surface coverages and assist in understanding the experimental results. Finally, at the optimal functionalization conditions, MoS$_2$ functionalized with Me-OED achieves a carrier density of $1.10 \pm 0.37 \times 10^{14}$ cm$^{-2}$, a record for the surface functionalization of MoS$_2$. Overall, our results establish that factors other than the molecule's redox properties impact molecular doping, and need to be carefully considered in the design of new OEDs.

**Results**

**Surface functionalization and carrier density change**

Me- and $^t$Bu-OED are neutral reductants that are stable and soluble in organic solvents under an inert atmosphere. $^t$Bu-OED was synthesized according to a previously published procedure,[24] while Me-OED is a novel compound (the synthesis and characterization is described in the Methods and Supplementary Information). The cyclic voltammograms and molecular structures of both OEDs are shown in Figure 1a. Me- and $^t$Bu-OED undergo two discrete one-electron transfer events at -0.91 V and -0.67 V, indicating they have identical redox potentials. Figure 1b shows the ground state structures of a Me- and $^t$Bu-OED molecule on monolayer MoS$_2$ as determined using DFT calculations (Details in Methods). The aromatic rings of both OEDs are nearly parallel to the basal plane of MoS$_2$, and Me-OED occupies a smaller surface area than $^t$Bu-OED on MoS$_2$, due to the smaller methyl substituents.

MoS$_2$ FETs were fabricated on monolayer MoS$_2$ flakes grown by chemical vapor deposition, and their transfer characteristics were measured before and after functionalization (Details in Methods). In an argon-filled glovebox10 mM Me- and $^t$Bu-OED solutions were drop-cast on MoS$_2$ FETs for 10 minutes and then washed with acetonitrile to remove any excess OED. The transfer characteristics of the functionalized MoS$_2$ devices were measured under an ambient atmosphere, as shown in Figure 1c. At a bias voltage of 1 V, pristine MoS$_2$ shows strong gate dependency with the expected n-type transport. After functionalization, the drain current ($I_D$) at zero gate voltage ($V_{GS}$) increased by three orders of magnitude for the Me- and $^t$Bu-OED functionalized MoS$_2$ FETs. $V_{GS}$ shifted to more negative values after functionalization, indicating that both OEDs are n-type dopants. The average 2D sheet carrier density ($n_{2D}$) in MoS$_2$ was obtained from the characteristic curves of several pristine and functionalized devices and is shown in Figure 1d (See Methods for calculation details). The average $n_{2D}$ is $3.48 \pm 0.28 \times 10^{11}$ cm$^{-2}$ for pristine MoS$_2$, $6.07 \pm 0.55 \times 10^{12}$ cm$^{-2}$ for $^t$Bu-OED functionalized MoS$_2$, and $2.89 \pm 0.73 \times 10^{13}$ cm$^{-2}$ for Me-OED functionalized MoS$_2$. Thus, despite the fact that the dopant molecules have the same redox potential, we observe that MoS$_2$ is doped more by Me-OED than by $^t$Bu-OED using the same functionalization conditions.

We next systematically varied the functionalization conditions for Me- and $^t$Bu-OED and measured the change in $n_{2D}$ of MoS$_2$ after functionalization. For both OEDs, we observe that a higher solution concentration or a longer exposure time leads to more doping, as evidenced by the dramatic increase in $I_D$ of MoS$_2$ functionalized with a 10 mM solution for 24 h compared to the case of a 0.1 mM solution for 10 minutes (Figure 2a). Figure 2b plots the average $n_{2D}$ of

MoS$_2$ for all the functionalization conditions we utilized. As expected, increasing the functionalization time and solution concentration resulted in increased $n_{2D}$. The only exception is using $^t$Bu-OED with 10 minutes of exposure time, where increasing the $^t$Bu-OED concentration from 0.1 to 10 mM did not change the $n_{2D}$. Overall, Me-OED is typically found to donate more electrons to MoS$_2$ than $^t$Bu-OED (see below for further explanation), and the MoS$_2$ functionalized with 10 mM Me-OED solution for 24 h achieves an average $n_{2D}$ of $1.10 \pm 0.37 \times 10^{14}$ cm$^{-2}$, the highest doping level reported for molecular doping to MoS$_2$.

**Molecular surface coverage and doping power of OEDs**

The molecular doping power is equal to the increase in MoS$_2$ carrier density from molecular functionalization ($\Delta n_{2D}$) divided by the number of molecules that donate charge. We previously found that X-Ray photoelectron spectroscopy (XPS) characterization only yields information about the total number of molecules on the MoS$_2$ surface, without consideration as to how the molecules are arranged and their doping power.[3] In contrast, AFM can provide reliable estimates of the surface coverage of molecules for lower concentrations,[3] allowing for the calculation of the doping power. Surface topographic images of MoS$_2$ flakes functionalized with 0.1 mM solutions of Me- and $^t$Bu-OED for 10 minutes were obtained with AFM. Figure 3 shows representative images of Me-OED functionalized MoS$_2$ (Figure 3a, b), $^t$Bu-OED functionalized MoS$_2$ (Figure 3d, e), and the height distributions for islands formed from each OED (Figure 3c, f). For the functionalized MoS$_2$, the surface roughness of the uncovered areas was comparable to that of pristine MoS$_2$ flakes, indicating that the uncovered areas are free of isolated molecules. The AFM images reveal that both OEDs tend to aggregate into islands. Me-OED forms discrete, round islands whereas $^t$Bu-OED forms islands that are bulky, interconnected, and chain-like. The average diameter of the Me-OED islands is 10.50 nm after deconvolving the AFM tip diameter, and there is a broad distribution of island heights (Figure 3c). The lack of clear step heights suggests that Me-OED molecules do not form discrete layers in which the aromatic ring of the molecule stacks on top of the $MoS_2$ surface or on top of each other. In contrast, the height distribution of the $^t$Bu-OED islands shows sharp peaks at 0.80 nm and 1.25 nm (Figure 3f). Considering the height of the $^t$Bu-OED molecule (0.45 nm) and the distance between the molecule and basal plane of $MoS_2$ (~0.20 nm), these values suggest formation of well-defined mono- and bi-layers of $^t$Bu-OED islands.

Analysis of the AFM images reveals that the average surface coverage is 40% for Me-OED and 56% for $^t$Bu-OED for functionalization conditions of 0.1 mM and 10 minutes. The average surface coverage increases to 82% for Me-OED and 57% for $^t$Bu-OED for functionalization conditions of 10 mM and 10 minutes (Supplementary Figure S1). We note that functionalization of $MoS_2$ with only the solvent (acetonitrile) yields an average surface coverage of 0.2%, thus the islands observed are not from the solvent molecules. The differences in island morphology and surface coverage between the two OEDs on $MoS_2$ indicate that the molecule-$MoS_2$ interactions must be distinct for the two OEDs due to their different sizes. This suggests that their doping powers to $MoS_2$ will be different despite their identical redox potentials.

The doping powers of Me- and $^t$Bu-OED were calculated using the surface coverage estimated by AFM, following our previous study on DMAP-OED molecules (see Methods).[3] The doping power for $^t$Bu-OED is estimated to be -0.11$e$ per molecule (where $e$ stands for the elementary charge) for both the 0.1 mM and 10 mM solutions for the exposure time of 10 minutes, assuming only the molecules closest to $MoS_2$ donate charge. The doping power of $^t$Bu-OED remains the same as the average $n_{2D}$ and the surface coverage did not change between the

0.1 mM and 10 mM concentrations for the same exposure time of 10 minutes. Only when the exposure time increased to 24 h, the average $n_{2D}$ and surface coverage increased for ᵗBu-OED. Thus, we hypothesize that the surface kinetics of ᵗBu-OED reaching a thermodynamic distribution on $MoS_2$ is slow. Essentially, an initial kinetic distribution is reached after 10 minutes and then the molecules slowly settle over the 24 hour period. At functionalization conditions of 10 mM solution for 10 minutes, Me-OED donates -0.22 to 0.44$e$ per molecule, depending on whether the molecule's aromatic ring is assumed to be tilted from or parallel with the basal plane of $MoS_2$ (see Methods). At functionalization conditions of 0.1 mM solution for 10 minutes, the estimated doping power of Me-OED was far below -0.11$e$ per molecules. At such low solution concentrations, we suspect that molecules are more prone to oxidation, which would lead to degradation of the molecules and weakening of their doping ability. This also explains why under these conditions ᵗBu-OED, which is likely less sensitive to oxidation due to its increased protection from the bulky ᵗBu-groups, is a better dopant.

**Density functional theory calculations for doping power of OEDs**

We performed DFT studies investigating Me- and ᵗBu-OED on monolayer $MoS_2$ under varying surface coverages to understand the interplay between molecular structure and doping power from first principles. We first consider monolayer and sub-monolayer molecular coverages on $MoS_2$. For both OEDs, the densest coverage simulated for a monolayer of molecules with their aromatic rings parallel to the $MoS_2$ surface is 1/12 (one molecule per 12 formula units of $MoS_2$; Figure 4a for the Me-OED case). We observe that the charge donation per molecule is similar for Me- and ᵗBu-OED for the same molecular coverages up to 1/12 (Figure 4c), which is expected due to the identical redox potentials for the two OEDs. The charge transfer per molecule decreases as the coverage increases, and at a coverage of 1/12, the charge transfer per molecule is computed to be -0.36$e$ and -0.38$e$ per molecule for Me- and $^t$Bu-OED, respectively. Higher monolayer coverages than 1/12 are possible if the aromatic rings can be tilted rather than parallel to the MoS$_2$ surface. For Me-OED, a monolayer of molecules with tilted aromatic rings is possible for coverages of 1/6 and 1/4.5 (Figure 4b for 1/6 coverage), and the doping power per molecule is computed to be -0.24$e$ for 1/6 coverage and -0.21$e$ for 1/4.5 coverage (Figure 4c). However, our DFT calculations find that a 1/6 coverage of $^t$Bu-OED with tilted molecules is unstable. Such a finding is expected because the bulky $^t$Bu groups are likely to hinder the formation of a densely packed monolayer.[25]

Our experimental results also show that bilayers of OED molecules on MoS$_2$ can be readily formed. The binding energies per molecule for the bilayers are just ~0.1-0.3 eV less than those for the monolayers (Table 1). For Me-OED, a bilayer at 1/12 surface coverage has the same number of molecules per MoS$_2$ units as a monolayer at 1/6 surface coverage. Comparing the energetics of these two systems, we find that Me-OED molecules prefer to form a monolayer with tilted aromatic rings rather than a bilayer with parallel aromatic rings. These predictions are consistent with the AFM height distributions (Figure 3c, f), which show clear step heights corresponding to mono- and bilayers for $^t$Bu-OED molecules, but a broad distribution of island heights for Me-OED. Our calculations show that the molecules in the top layer away from the MoS$_2$ surface do not contribute to charge donation, similar to our earlier findings for DMAP-OED.[3]

**Comparison between experimental results and DFT calculations**

The calculated doping powers of densely covered Me-OED (-0.21$e$ to -0.36$e$ per molecule at 1/4.5 and 1/6 surface coverage) agree well with the experimental values (-0.22$e$ to -0.44$e$), and the molecule's orientation in relation to MoS$_2$ basal plane impacts the doping power of Me-OED. However, for $^t$Bu-OED, the calculated doping power per molecule (-0.38$e$) is much larger than the experimental value (-0.11$e$). Our current hypothesis for this discrepancy is that AFM may overestimate the number of $^t$Bu-OED molecules on the surface, which would result in an underestimate of the doping power. We assume that the $^t$Bu-OED islands observed by AFM take the molecular arrangement of the densely packed 1/12 surface coverage. However, this molecular arrangement might not be reached for $^t$Bu-OED for the exposure time of 10 minutes when the surface kinetics of $^t$Bu-OED on MoS$_2$ is slow. This reasoning is based on our observation that the surface coverage for $^t$Bu-OED did not change with solution concentration when the settling time was 10 minutes. Thus, there may be fewer $^t$Bu-OED molecules in the observed molecular islands than what we estimate. The lateral resolution of AFM is insufficient to image the molecular arrangements.

Having calculated the molecular doping powers of Me- and $^t$Bu-OED to MoS$_2$ at varying surface coverages, we also computed the average $n_{2D}$ of functionalized MoS$_2$ to compare with the experimentally determined $n_{2D}$. The computed $n_{2D}$ is $5.2 \times 10^{13} cm^{-2}$ for Me-OED at 1/4.5 coverage and $3.5 \times 10^{13} cm^{-2}$ for $^t$Bu-OED at 1/12 coverage (Figure 4d), which are in the same order of magnitude as the experimental values (Figure 2b). However, for the optimal functionalization conditions (10 mM solution, 24 h), the computed $n_{2D}$ is almost half the experimental $n_{2D}$. With a settling time of 24 h, we hypothesize that OED molecules might intercalate between the MoS$_2$ and the substrate, functionalizing both surfaces of the MoS$_2$ monolayer. We simulate this scenario for a 1/4.5 surface coverage of Me-OED and a 1/12 surface coverage for ᵗBu-OED (the densest monolayer coverage). Figure 5a and 5b show the corresponding relaxed geometries for the Me- and ᵗBu-OED cases, respectively. We find that adsorption on both sides of MoS₂, instead of one side, does not change the binding energies per molecule significantly, while the corresponding electron doping density is approximately doubled. As a result, the highest $n_{2D}$ computed are $1.04 \times 10^{14} cm^{-2}$ for Me-OED functionalized MoS₂, and $7.0 \times 10^{13} cm^{-2}$ for ᵗBu-OED functionalized MoS₂ (Figure 5c). These numbers are in good agreement with the experimental results ($1.10 \pm 0.37 \times 10^{14} cm^{-2}$ for Me-OED and $4.83 \pm 0.90 \times 10^{13} cm^{-2}$ for ᵗBu-OED for under the optimal functionalization conditions; Figure 2b).

**Conclusion**

Our results demonstrate that despite having the same redox potential, the doping powers of Me- and ᵗBu-OED vary due to differences in their size. This affects their interactions with MoS₂, which is reflected in the distinct morphologies and height distributions of the Me- and ᵗBu-OED islands. We observe that the steric properties of the molecule affect the binding energy between the molecules and MoS₂, determining whether the molecules agglomerate laterally or vertically. Further, molecular size impacts the molecular packing efficiency, an important parameter for the total charge donation to MoS₂. The stronger doping power of Me-OED compared to ᵗBu-OED is attributed to Me-OED's smaller size, which allows for better packing efficiency, thus maximizing the total charge transferred from the molecules to MoS₂. DFT calculations of the molecular doping power and achievable carrier density agree well with the experimental results, and demonstrate that the highest carrier density can be achieved via double-surface functionalization. Thus, our findings demonstrate that the molecule with the highest redox potential in isolation may not lead to the highest achievable carrier density in a 2D material, even at the optimal functionalization conditions. Powerful OED dopants should possess a maximally negative redox potential for maximum charge donation per molecule and be small in size so a maximum number of molecules can come in contact with the 2D surface. In conclusion, our results clearly demonstrate that properly tuning the molecular properties beyond the redox potential is necessary to develop powerful molecular dopants.

**Methods**

**Synthesis and characterization of Me- and ᵗBu-OED.**

*Reagents used in the synthesis of Me-OED and ᵗBu-OED*

Acetonitrile ($CH_3CN$) was purchased from Honeywell (Cat. No. CS017-56) and used without further purification. Pentane was dried via passage through a column of activated alumina on an Inert Technologies PureSolv MD7 solvent purification system and subsequently stored under dinitrogen. Toluene-$d_8$ was degassed via three consecutive freeze-pump-thaw cycles and dried by passage through a short column of neutral activated alumina and stored under an $N_2$ atmosphere. Neutral alumina was activated by heating at 250 °C *in vacuo* overnight. [Me-OED$^{2+}$][Br$^-$]$_2$ and [ᵗBu-OED$^{2+}$][Br$^-$]$_2$ were synthesized according to literature procedures.[24, 26]

*General procedure*: Inside of a glovebox under an $N_2$ atmosphere, a 2 dram scintillation vial was charged with a magnetic stir bar, [R-OED$^{2+}$][Br$^-$]$_2$ (R = Me, ᵗBu) (1 equiv.), Mg$^0$ (3 equiv.), and dry degassed $CH_3CN$ (0.3 M). The reaction was stirred at room temperature for 4 hours, during which the reaction turned deep purple. The volatiles were removed *in vacuo* and pentane (0.15 M) was added. The reaction was filtered under $N_2$ and the volatiles were removed to afford the desired product. The products were characterized using NMR spectroscopy (Supplementary Figure S2 and S3) and high resolution mass spectrometry (HRMS) (Supplementary Figure S4).

*tBu-OED*: 0.18 g, (65%). The NMR data are consistent with a previous literature report.[24]

*Me-OED*: 0.14 g, (59%). $^1$H NMR (500 MHz, toluene-$d_8$, -70 °C) δ 5.88 (s, 2H), 5.31 (s, 2H), 4.75 (s, 2H), 2.73 (s, 4H), 1.66 (s, 6H), 0.62 (s, 2H). $^{13}$C{$^1$H} NMR (125 MHz, toluene-$d_8$, -70 °C) δ 122.55, 122.15, 107.12, 56.31, 26.28, 26.18. (HRMS) TOF MS ES$^+$ (m/z) [M]$^+$ calculated for [$C_{15}H_{18}N_2$]$^+$ 226.1465; found 226.1454.

**CVD synthesis of MoS$_2$.**

Monolayer MoS$_2$ flakes were synthesized on SiO$_2$/Si substrates in a 1-inch quartz tube furnace. 0.4 mg of MoO$_3$ powder was placed in a quartz crucible at the center of the furnace, and 200 mg of sulfur powder was placed upstream in an alumina crucible liner, with 17 cm separating the precursors.[27] A 285 nm SiO$_2$/Si substrate was treated with piranha solution (3:1 H$_2$SO$_4$ : H$_2$O$_2$) for at least 3 hours, and subsequently rinsed with deionized water and dried using a stream of N$_2$ gas. The substrate was then treated with a single drop (1.5 μL) of 100 μM perylene-3,4,9,10 tetracarboxylic acid tetrapotassium salt (PTAS). After drying the PTAS-treated substrate on a hot plate in air, the substrate was placed face down on the MoO$_3$ crucible. The quartz tube was purged several times with ultra-high purity argon gas to ensure no residual oxygen was present. The furnace temperature was ramped to 850 °C, and then kept at 850 °C for 15 min, while flowing argon at 10 sccm. After the reaction, the furnace was naturally cooled to 580 °C, and then the lid was opened to accelerate the cooling to room temperature. Raman spectroscopy, photoluminescence (PL), and AFM were performed to confirm the formation of monolayer MoS₂ flakes.

**Device fabrication and characterization.**

To avoid gate leakage, the as-grown MoS₂ monolayers were transferred to a fresh SiO₂ (285 nm) / Si (p⁺) substrate, which served as back-gate for field-effect devices. For the transfer of MoS₂, 950 PMMA A4 (MicroChem) was spin-coated on the growth SiO₂ substrates containing MoS₂ flakes and baked at 120 °C for 5 minutes on a hot plate. The PMMA / MoS₂ film was released from the SiO₂ substrate by floating the sample in 2M KOH at 65 °C for one hour. The film was then rinsed with deionized water several times and transferred to a fresh SiO₂ / Si substrate. After drying the sample on a hot plate at 40 °C for 40 minutes, the sample was kept in acetone overnight to remove the PMMA. Electron beam lithography was used to pattern the metal contacts followed by e-beam evaporation of Ti (10 nm) / Au (80 nm) and lift-off in acetone overnight. Electrical measurements were performed on these devices before and after molecular functionalization. The electrical characteristics of the devices were measured in air using a semiconductor device analyzer (Agilent Technologies B1500A).

**Carrier density calculations.**

From the FET transfer curves, the 2D sheet carrier density in MoS₂ was calculated[7] by $n_{2D} = (I_D L)/(qWV_{DS}\mu)$, where $I_D$ is the drain current at the zero gate voltage; $L$ and $W$ are the length and width of the channel, respectively; $q$ is the electron charge, and $\mu$ is the field-effect mobility at $V_{DS}$ = 1 V. The mobility was calculated as $\mu = (\partial I_D/\partial V_{GS}) L/(V_{DS} C_{ox} W)$, where $(\partial I_D/\partial V_{GS})$ is the maximum transconductance, and $C_{ox}$ is the gate capacitance of $1.2 \times 10^{-8}$ F/cm² for 285 nm thick SiO$_2$ estimated based on the parallel-plate model. Numerous MoS$_2$ devices were measured to ensure reproducibility of the observed transfer curves and to obtain the average $n_{2D}$ and the standard error, which represent the error bars (Fig. 1d and 2b).

**AFM characterization and surface coverage calculations.**

AFM images were acquired with a Cypher ES Environmental AFM System (Asylum Research Oxford Instruments) in tapping mode using the FS-1500AuD (Asylum Research Oxford Instruments) cantilever at a scan rate between 4.88 and 7.81 Hz. All imaging was performed under ambient conditions. To determine the surface coverage of Me- and $^t$Bu-OED on functionalized flakes, hundreds of AFM images were collected from different areas of multiple pristine and functionalized MoS$_2$ flakes with scan sizes suitable to view the entire flake, as well as windows of 1.5 μm, 1 μm, 500 nm, 250 nm, and 50 nm. The AFM images were processed using Gwyddion and ImageJ to extract the height and size of the molecular islands, as well as the average areal surface coverage.

To estimate the surface coverage of molecules, AFM images were processed using ImageJ to set a background threshold, perform particle analysis, and compute the percent surface coverage (*SC*). For each molecular functionalization condition, the same analysis was performed for at least 10 samples and the results were averaged. For the calculation of the number of molecules in the MoS$_2$ FET channel, we assume that the molecules are densely packed with a fixed buffer zone between adjacent molecules. This is equivalent to the DFT surface coverage of 1/12 for $^t$Bu-OED, corresponding to a molecular surface area of 106.067 Å$^2$ per molecule. For Me-OED the two physical configurations, tilted and parallel, were considered. The tilted configuration (i.e. 1/6 coverage) corresponds to a surface area of 53.033 Å$^2$ per molecule and the flat configuration (i.e. 1/12 coverage) corresponds to a surface area of 106.067 Å² per molecule. Thus, the total number of molecules ($N_M$) with surface area ($SA$) in the MoS$_2$ FET channel with length $L$ and width $W$ is estimated to be $N_M = \frac{SC \times L \times W}{SA}$.

**DFT calculations.**

Spin-polarized DFT calculations were performed for both OED functionalized MoS$_2$ monolayers using the SIESTA code[28], with the PBE generalized gradient approximation[29] for the exchange-correlation functional and Grimme's pair-wise DFT-D2 parameters[30] to account for the van der Waals interactions. Troullier-Martins type norm-conserving pseudopotentials were used, together with a double-zeta plus polarization basis sets. The Brillouin zone was sampled by a k-point mesh that is equivalent to a 27×27×1 k-grid or better for the primitive MoS$_2$ unit cell. A mesh cutoff energy of 300 Ry was used to obtain the electronic wavefunctions and charge densities. Atomic positions were relaxed until the forces were smaller than 0.02 eV/Å. For calculating the doping power of organic OEDs, Bader charge analysis[31] was performed with a finer FFT mesh of at least 240×240×960. The binding energy per molecule is computed as $E_{BE} = (E_{MoS2} + E_{OED,gas-phase} - E_{OED-MoS2})/N_{mol}$, where $N_{mol}$ is the number of molecules in the supercell.

**Acknowledgements**

The synthesis of MoS$_2$ flakes was supported by a CAREER award from the NSF (1749742). S. N. R. acknowledges the Ford Foundation for a graduate student fellowship, while J. B. C. thanks the NSF for a graduate research fellowship. N. H. acknowledges support from a seed grant from the Center for Research on Interface Structures and Phenomena at Yale University. We thank Dr. Fabian Menges for assistance with high-resolution mass spectrometry and Dr. Yiren Zhong and Dr. Hailiang Wang for their help with surface functionalization. Material characterizations were carried out at shared facilities including the Yale West Campus Materials Characterization Core, the Yale Institute for Nanoscience and Quantum Engineering, and the School of Engineering Cleanroom. S. Y. Q. and Y. C. acknowledge funding from the Grant MOE2016-T2-2-132 from the Ministry of Education, Singapore, and support from the Singapore National Research Foundation, Prime Minister's Office, under its medium-sized centre program. Computations were performed on the NUS Graphene Research Centre cluster and National Supercomputing Centre Singapore.

**Competing interests**

The authors declare no competing interests.

**Supporting information**

Supplementary information is available for this manuscript.

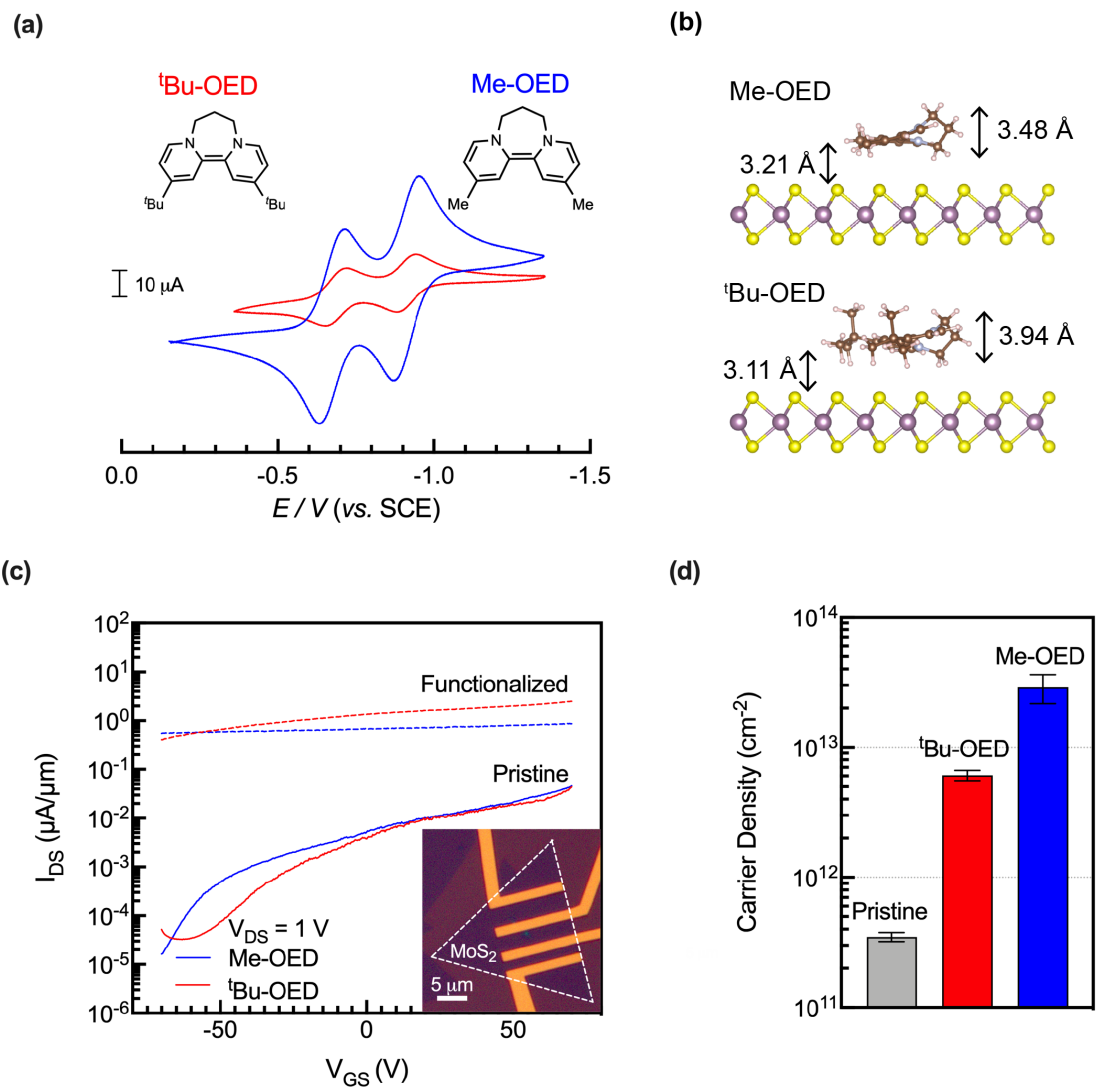

**Figure 1. Cyclic voltammograms (CVs) of the molecular dopants and transfer characteristics of MoS$_2$ transistors. a)** Chemical structures and CV scans of Me- and $^t$Bu-OED, indicating that the molecules have identical redox potentials. **b)** Atomic structure of isolated Me-OED (top) and $^t$Bu-OED (bottom) molecules on MoS$_2$ from DFT calculations. **c)** Representative $I_D$-$V_{GS}$ transfer curves of pristine MoS$_2$ FETs and after functionalization with either 10 mM Me- or $^t$Bu-OED solution for 10 minutes. The bias voltage is 1 V and the channel lengths are 3 μm. Inset: Optical image of a pristine MoS$_2$ FET with varying channel lengths. **d)** Average carrier

density at $V_{GS} = 0$ V for pristine (gray), $^t$Bu-OED (red), and Me-OED (blue) functionalized MoS$_2$ FETs prepared using a 10 mM solution and a 10 minute exposure time. The error bars represent the standard error obtained from measuring 5 to 8 MoS$_2$ devices.

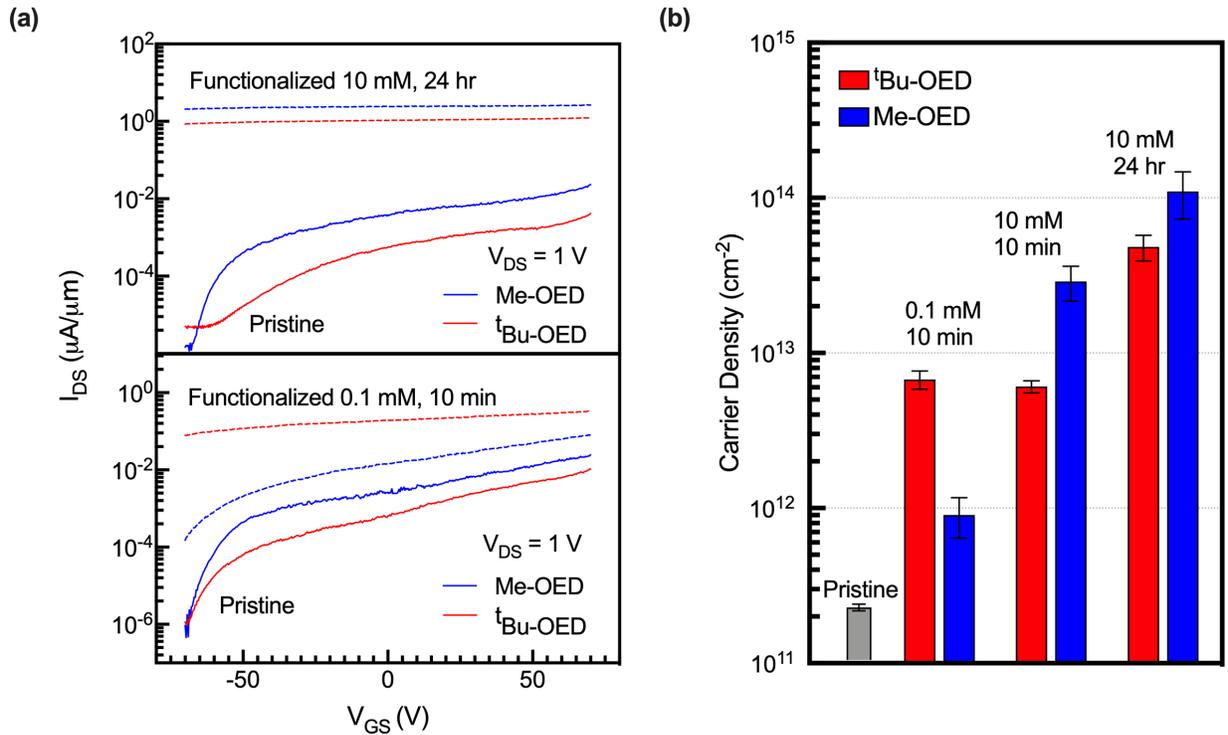

**Figure 2. Systematic functionalization of MoS$_2$ with Me- and $^t$Bu-OED. a)** Representative $I_D$-$V_{GS}$ transfer curves of MoS$_2$ FETs functionalized with Me- and $^t$Bu-OED for two different treatment conditions: 10 mM, 24 hours (top) and 0.1 mM, 10 minutes (bottom). The bias voltage is 1 V and the channel length of the devices is 3 μm. **b)** Systematic increase of the carrier density of MoS$_2$ functionalized with Me- and $^t$Bu-OED at $V_{GS} = 0$ V with increasing solution concentration and functionalization time. The error bars represent the standard error obtained from measuring 5 to 8 MoS$_2$ devices.

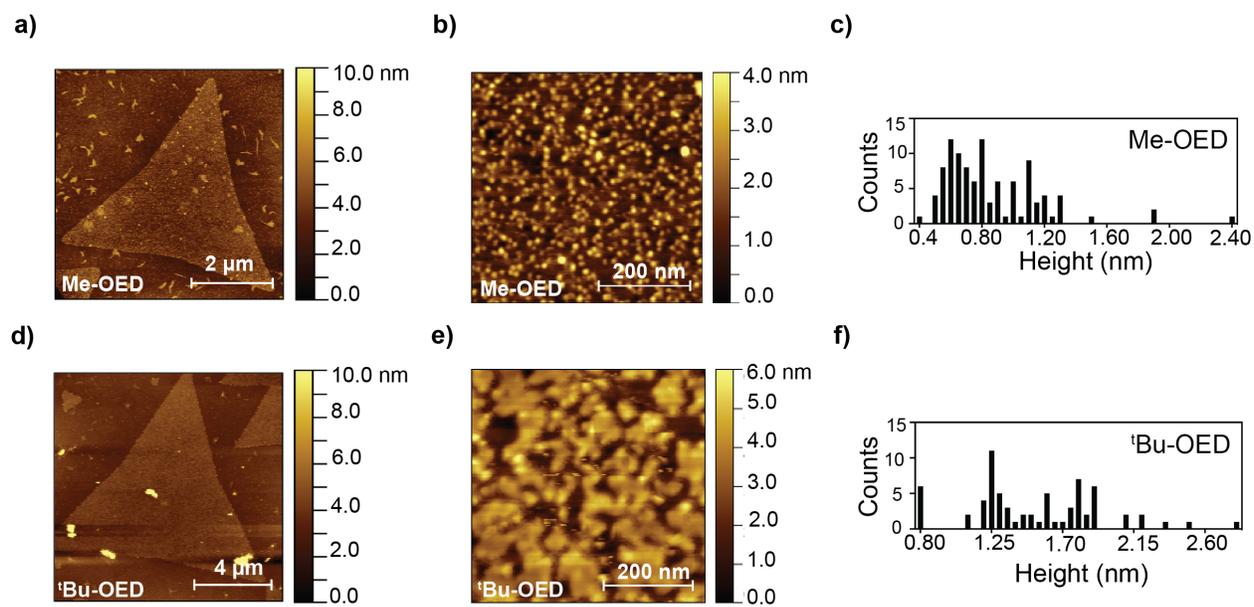

**Figure 3. AFM analysis of MoS$_2$ functionalized with Me- and $^t$Bu-OED.** AFM images of MoS$_2$ functionalized with 0.1 mM Me-OED for 10 minutes **(a-b)** and with 0.1 mM $^t$Bu-OED for 10 minutes **(d-e)**. Me-OED tends to aggregate into discrete, round islands, whereas $^t$Bu-OED aggregates into bulky, chain-like islands. Height distributions of the molecular islands formed on MoS$_2$ after functionalization with Me-OED **(c)** and $^t$Bu-OED **(f)**.

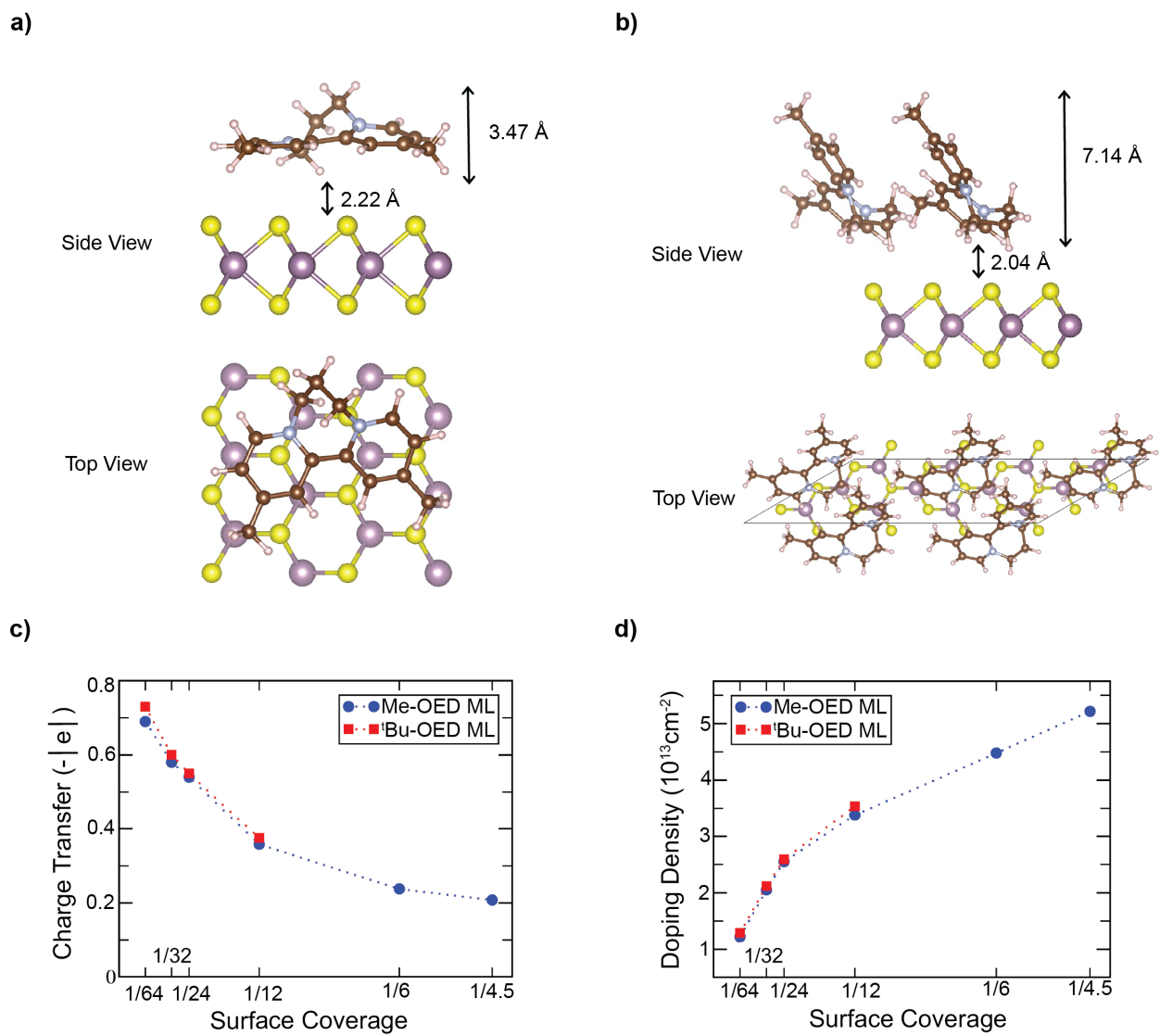

**Figure 4. Molecular arrangements and doping powers as a function of surface coverage. (a, b)** Monolayer arrangements of Me-OED molecules on MoS$_2$: side view (top panel) and top view (bottom panel) at (a) 1/12 coverage and (b) 1/6 coverage. **(c)** Charge transfer per molecule as a function of surface coverage. **(d)** Total electron doping density of monolayer OEDs on MoS$_2$ as predicted by DFT. The OED surface coverages are in units of one molecule per number of formula units of MoS$_2$ (e.g. a coverage of 1/12 refers to one molecule per 12 formula units of MoS$_2$). Dotted lines are guides for the eye.

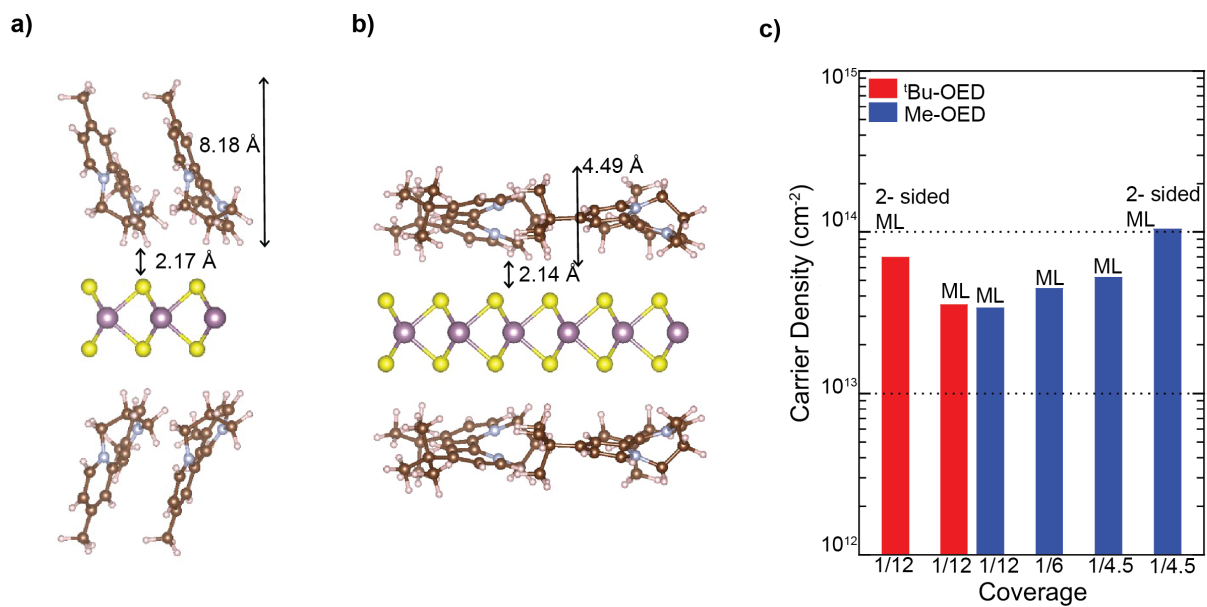

**Figure 5. Double surface-functionalization with OEDs. (a)** Side view of monolayers of Me-OED with tilted rings covering both sides of MoS$_2$ at 1/4.5 surface coverage (i.e., two molecules cover a 3x3 supercell of MoS$_2$). **(b)** Side view of monolayers of $^t$Bu-OED covering both sides of MoS$_2$ at 1/12 surface coverage. **(c)** Summary of DFT-calculated total electron doping densities for both OEDs at the denser coverage regimes for monolayer coverage on a top surface of MoS$_2$ (labeled 'ML') or both surfaces of MoS$_2$ (labeled '2-sided ML').

**Table 1.** DFT-calculated binding energies per molecule for monolayer and bilayer OED molecules on MoS$_2$.

| Binding energy (eV) | Me-OED @1/12 surface coverage | Me-OED @1/6 surface coverage | $^t$Bu-OED @ 1/12 surface coverage |
|---|---|---|---|
| Monolayer | 1.74 | 1.79 | 2.07 |
| Bilayer (on top of MoS$_2$) | 1.44 | 1.63 | 1.88 |
| Bilayer (molecule on top rotated by 90 degrees) | 1.52 | - | 1.89 |